\documentclass[12pt, oneside]{article}   	
\usepackage{geometry}                		
\usepackage{graphicx}					
\usepackage{amssymb}
\usepackage{color}
\usepackage{authblk}
\usepackage{setspace}
\usepackage{draftwatermark}
\SetWatermarkScale{2}
\SetWatermarkLightness{0.5}

\title{\textbf{DRAFT -- Under Review} \\ 
Almost Politically Acceptable Criminal Justice Risk Assessment}
\author[1,2]{Richard Berk}
\author[3]{Ayya A. Elzarka}
\affil[1] {Department of Criminology, University of Pennsylvania}
\affil[2] {Department of Statistics, University of Pennsylvania}
\affil[3] {Google LLC, Mountain View, CA.} 

\begin{document}
\maketitle

\begin{abstract}
\noindent \textbf{Research Summary} \\
In criminal justice risk forecasting, one can prove that it is impossible to optimize accuracy and fairness at the same time. One can also prove that it is impossible optimize at once all of the usual group definitions of fairness. In the policy arena, one is left with tradeoffs about which many stakeholders will adamantly disagree. In this paper, we offer a different approach. We do not seek perfectly accurate and fair risk assessments. We seek politically acceptable risk assessments. We describe and apply to data on 300,000 offenders a machine learning approach that responds to many of the most visible charges of ``racial bias.'' Regardless of whether such claims are true, we adjust our procedures to compensate. We begin by training the algorithm on White offenders only and computing risk with test data separately for White offenders and Black offenders. Thus, the fitted algorithm structure is exactly the same for both groups; the algorithm treats all offenders as if they are White. But because White and Black offenders can bring different predictors distributions to the white-trained algorithm, we provide additional adjustments if needed. 

\noindent \textbf{Policy Implications} \\
Insofar are conventional machine learning procedures do not produce accuracy and fairness that some stakeholders require, it is possible to alter conventional practice to respond explicitly to many salient stakeholder claims even if they are unsupported by the facts. The results can be a politically acceptable risk assessment tools.  

 
\noindent Direct correspondence to Richard Berk, Department of Criminology, McNeil Hall, University of Pennsylvania, Philadelphia, PA. (e-mail: berkr@sas.upenn.edu)
\end{abstract}

\section*{Introduction}

The literature on fairness for algorithmic, criminal justice risk assessments is large and continues to grow (Kleinberg et al., 2017; Berk et al., 2018; Kearns et al., 2018; Hug, 2019; Mayson, 2019). Many of the issues are complex. There are inherent tradeoffs between different kinds of fairness and between fairness and accuracy. Despite well-intended aspirations, you can't have it all. Various proposed technical solutions typically select one or two kinds of fairness for which they can provide a ``fair'' algorithm. Other forms of fairness and the fairness tradeoffs are ignored. Reductions in accuracy are commonly an afterthought. There is, moreover, no single, dominant kind of fairness. Different stakeholders stubbornly can hold different and legitimate conceptions of fairness.

There is no clear resolution likely in the near term. Meanwhile, criminal justice decisions will be made for many thousands of offenders. Various forms of risk assessments commonly will inform those decisions. In this paper, we propose a fallback position that might be applied immediately to the construction of risk assessments. Rather than fair risk assessment, we offer an approach that might be called politically acceptable risk assessment.

\section*{Politically Acceptable Risk Assessment}

Criticisms of the role of race in the American criminal justice system are often well-founded. But, just as for fairness itself, the issues are complicated. Apprehensions by the criminal justice system are often the last stop on a train that left the station early in an offender's childhood. ``Mass incarceration,'' for example, is a product of many factors, including widespread exposure to violence early in life (Bloom, 2014) and many different kinds of disadvantage. 

Properly understood, risk algorithms are meant to help inform criminal justice decisions to make them more accurate and more fair; the benchmark is current practice. Algorithms will surely make some forecasting errors and will surely produce some unfairness, but the aspiration is improved accuracy and fairness. Initial evidence indicates that both can materialize in real settings (Berk, 2018). 

\begin{figure}[htbp]
\begin{center}
\includegraphics[width=4in]{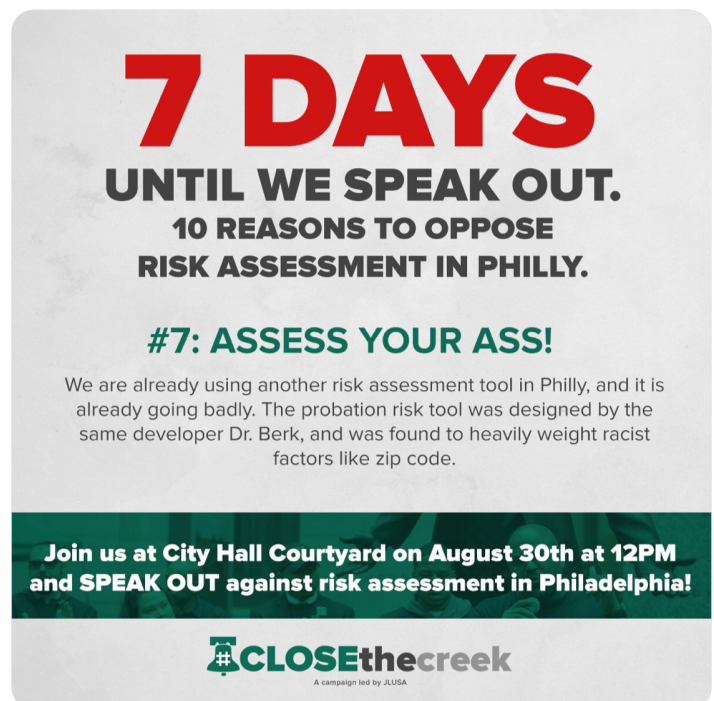}
\caption{Poster From Anti-Risk Assessment Rally}
\label{fig:rally}
\end{center}
\end{figure}

Nevertheless, risk algorithms can elicit a primal response too often supported by deep misunderstandings. Figure~\ref{fig:rally} illustrates one instance. The existing risk assessment tool had in fact been a great success, did not use zip code as a predictor, and justified far less intrusive probation supervision for low risk offenders, who were disproportionately African-American (Berk et al., 2010). 

Given deep and and often unbending opposition from some stakeholders and the impossibility of a perfectly accurate and absolutely fair risk instrument, it might make sense to focus instead on algorithmic risk instruments that are \textit{politically} acceptable. To simplify the exposition, we will proceed assuming offenders are either African-American or White. Similar issues can arise for other racial and ethnic groups, and our approach easily generalizes. 

Claims often are made that criminal justice decisions are tainted by ``white privilege'' (Harcourt, 2007; Star, 2014). One might infer, therefore, that if Black offenders were sentenced in the same manner as White offenders, Black offenders would demonstrably benefit, and a form of equality would result. A practical response in the algorithmic world would be to train a risk algorithm solely on White offenders and then compute from test data the risks for Black and White offenders separately using the white-trained instrument. Black and Whites jointly would benefit from white privilege. We start there after laying some conceptual and empirical groundwork.\footnote
{
More specifically, a machine learning algorithm such as random forests or stochastic gradient boosting could be use to fit data only from white offenders. With the trained algorithmic results in hand, risk forecasts would be computed separately using test data for White offenders and Black offenders. This approach is related to procedures used to partition wage differentials by gender (Binder, 1973; Oaxaca, 1973), which was invented in demography about 20 years earlier (Kitagowa, 1955).
}  

\section*{Data and Statistical Methods}

We have a dataset of 300,000 offenders from a large metropolitan area arraigned very soon after an arrest. 67\% of the offenders are African-American, and 32\% of the offenders are white. We use an outcome variable with three classes: no arrest after an arraignment release, an arrest for a non-violent crime after an arraignment release, and an arrest for a violent crime after an arraignment release.\footnote
{
These releases assume a return for a subsequent court appearance. 
} 
These are the three classes for which risk will be computed. 

After release, 57\% of the offenders had no arrest, 33\% had an arrest for a non-violent crime, and 10\% had an arrest for a violent crime. The corresponding figures for Whites alone are 58\%, 35\%, and 7\%. For Blacks, the corresponding figures are 56\%, 32\% and 11\%. Perhaps the most notable difference by race is the likelihood of arrests for a violent crimes. Some might see this as an important difference in base rates between White and Black offenders.\footnote
{
The adjectives``non-violent'' and ``violent'' quite accurately describe the vast majority the crimes in their respective categories. But there are a few kinds of crime in the violent crime category that some criminal justice stakeholders wanted to include because of heightened public concerns, even if not violent in themselves. The most common examples were ``sex crimes'' such as child pornography, which exploits children for sexual stimulation. No assault need be involved. This decision may be re-visited in future discussions among stakeholders.
}

These comparisons can imply differential accuracy in current practice. Presumably, a magistrate would be very reluctant to release an offender who would be re-arrested for a violent crime. Yet, that apparently is what happens 7\% of the time for White offenders and 11\% of the time for Black offenders. More such mistakes are made for Black offenders, and because violent crimes typically are intra-racial, these mistakes disproportionately affect Black victims of violent crime.\footnote
{
Because of the very large number of offenders arraigned in a give year, the difference between 7\% and 11\% in current practice, translates into several thousand more victims of violent crime who will be disproportionately African-American.
}

The predictors routinely available in machine readable form included several biographical attributes (e.g., gender), information extracted from rap sheets (e.g., the number of prior arrests for burglary), and the immediate criminal charges alleged by arresting police officers prior to arraignment (e.g., three charges of aggravated assault)  At an arraignment, available information usually is very limited. 

We anticipated concerns about predictors derived from arrests recorded on rap sheets. One common complaint is that Black offenders often have longer rap sheets because of police ``bias.'' Those longer records, in turn, lead to unfavorable risk assessments. Police racial animus is commonly cited to explain arrests that are unjustified. A related claim is that Black neighborhoods are ``over-policed,'' even if the over-policing is well-intended: more police, more arrests. 

We take no position on these issues in this paper but observe that there are no doubt some police officers who unfairly target Black offenders. The question is whether such practices are frequent enough to impact arrest statistics in a significant manner. With respect to over-policing in major metropolitan areas, police deployments are dominated by calls for service (911 calls), and citizens in disadvantaged neighborhoods disproportionately call the police. A higher density of police can be a consequence of these calls. One would be hard pressed to see that higher density as police exercising racial animus, although longer rap sheets for Black offenders could result.

With the intent of implementing a politically acceptable risk assessment, we discarded all predictors derived from prior arrests for less serious crimes. These are crimes for which police might exercise substantial discretion (e.g., loitering).  Such discretion is said by some to introduce unjustified racial disparities. We also excluded all priors for arrests as a juvenile. We retained priors for arrests as an adults that were likely to be charged as felonies, often associated with violence and identifiable victims.  In the interest of equality, all of predictor exclusions and retentions were made for Black offenders and White offenders alike. 

Training data and test data were constructed as disjoint random splits of equal size. We planned to use stochastic gradient boosting, which is vulnerable to overfitting. Therefore, valid test data were essential. In addition, we anticipated subsetting the data by race. Implementing both  kinds of data partitioning required beginning with a large number of observations. Because the full dataset included 300,000 observations, each of the analyses to follow was undertaken with at least several thousand cases.

Stochastic gradient boosting (Hastie et al., 2009: Section 10.10), implemented in R as \textit{XGBoost}, was used for the analysis. The target cost ratios were to be the same for all classification errors. By weighting the data, all empirical cost ratios were, as intended, approximately 1 to 1. The 1 to 1 target cost ratios were a provisional decision made by criminal justice officials responsible for arraignments, and the cost ratios to be used in practice are still to be determined. Determining appropriate cost ratios is a very important component of any classification enterprise, but they are peripheral to the issues addressed in this paper. 

\section*{Results}

To set the context, we first applied stochastic gradient boosting to \textit{the entire training dataset} and then constructed confusion tables from test data separately for White offenders and Black offenders. Table~\ref{tab:regwhites} is the confusion table for Whites. 54\% of the White offenders were predicted not to be re-arrested after an arraignment release. 33\% were predicted to be re-arrested for a non-violent crime. 12\% were predicted to be re-arrested for a violent crime.\footnote
{
All of offenders were released with the expectation that they would be returning for a later court appearance. Charges were not dropped at the arraignment.
}

\begin{table}[htp]
\footnotesize
\caption{Stochastic Gradient Boosting Confusion Table from Test Data for White Offenders Using the Conventionally Trained Algorithm: 54\% Predicted No Arrest, 33\% Predicted Non-violent Arrest, 12\% Predicted Violent Arrest}
\begin{center}
\begin{tabular}{|c|c|c|c|c|}
\hline \hline
Observed & No Arrest  & Non-Violent Arrest & Violent Arrest & Classification \\
Outcomes & Predicted &  Predicted & Predicted &  Error \\ \hline
No Arrest  & 17877 & 6848 & 2535 & .34 \\
Non-Violent Arrest & 6454 & 7593 & 2062 & .53 \\
Violent Arrest & 1859 & 1779 & 1234 & .75 \\
\hline
Prediction Error & .32 & .53 & .79 & \\
\hline \hline 
\end{tabular}
\end{center}
\label{tab:regwhites}
\end{table}

Classification error is not especially relevant in this policy setting. It assumes that each outcome is known and then computes the fraction of cases that the algorithm misclassifies. When a decision needs to be made about a particular offender, the outcome is not known. On cannot tell in which row of the confusion table the offender belongs.\footnote
{
When there are more than two outcome classes, the terms ``false positive'' and ``false negative'' make no sense. For each actual outcome class, there can be two or more ways to misclassify. And with more than two classes, if one is called a ``positive'' and one is called a ``negative,'' what are the other classes to be called? As far as we know, there are no naming conventions that have addressed this problem.
}
For example, when the observed outcome is no arrest, that outcome is misclassified 34\% of the time. When the observed outcome is an arrest for a violent crime, that outcome is misclassified 75\% of the time. But which applies to any given offender when a decision is to release or detain? 

The difference in the two figures is a routine disparity that can result from unbalanced base rate distributions. It is usually more difficult to correctly classify outcomes that are substantially less common. If desired, one can respond by weighting the less common cases relatively more heavily, but that option was precluded here because the provisional 1 to 1 cost ratios would not longer hold.

Far more important for policy is forecasting accuracy when the outcome is \textit{not} known. With training and test data, generalization error can be estimated.\footnote
{
For categorical outcomes, the definition of generalization error depends on whether the fitted value is a class or a class probability (Hastie et al., 2009: 221). We are applying the definition for fitted classes. For a 0-1 loss, the usual MSE becomes the proportion forecasted incorrectly. 
}
From Table~\ref{tab:regwhites}, a forecast of no arrest is wrong 32\% of the time. A forecast of an arrest a non-violent crime is wrong 53\% of the time. A forecast of an arrest for a violent crime is wrong 79\% of the time. With different cost ratios it would be possible to do somewhat better, but there is still a demonstrable improvement. Applying a Bayes classifier to the marginal distribution of the outcome, one would forecast no arrest using none of the predictors and be wrong 43\% of the time. If no arrest is forecasted from Table~\ref{tab:regwhites}, the estimate of generalization drops to 32\%. That is a reduction in generalization error of about a quarter compared to current practice.\footnote
{
Given the large number of offenders arraigned in a given year, this could translate into thousands fewer crime victims. 
}

\begin{table}[htp]
\footnotesize
\caption{Stochastic Gradient Boosting Confusion Table from Test Data for Black Offenders Using the Conventionally Trained Algorithm: 55\% Predicted No Arrest, 33\% Predicted Non-violent Arrest, 12\% Predicted Violent Arrest}
\begin{center}
\begin{tabular}{|c|c|c|c|c|}
\hline \hline
Observed & No Arrest  & Non-Violent Arrest & Violent Arrest & Classification \\
Outcomes & Predicted &  Predicted & Predicted &  Error \\ \hline
No Arrest  & 38178 & 14301 & 5098 & .34 \\
Non-Violent Arrest & 13346 & 15749 & 4231 & .54 \\
Violent Arrest & 3792 & 3965 & 2817 & .74 \\
\hline
Prediction Error & .31 & .54 & .77 & \\
\hline \hline 
\end{tabular}
\end{center}
\label{tab:regblacks}
\end{table}

Table~\ref{tab:regblacks} is the confusion table for Black offenders constructed from test data.  55\% of the Black offenders were predicted to not be re-arrested after an arraignment release. 33\% were predicted to be re-arrested for a non-violent crime. 12\% were predicted to be re-arrested for a violent crime. The figures are virtually the same as those for White offenders. Black offenders are seen as no more or no less risky than White offenders. The figures for classification error and prediction error  are also nearly the same for Black offenders and White offenders.

But some caution is necessary. The near equivalences should not be taken too literally. Even if one treats these very large samples as populations, there is some imprecision in all of the proportions computed from the confusion tables. The main problem is that tuning machine learning algorithms is an approximation enterprise. Small differences in tuning parameter values can sometimes translate into variation of several percentage points when confusion tables are constructed and proportions computed. Nevertheless, it would be difficult to make a strong case for racial unfairness of these results.

Even though the risk assessments for Blacks and Whites were undertaken with the exact same trained algorithm, different offenders bring different predictor distributions to the test data, which can produce disparities between the two confusion tables depending on the way the algorithm is tuned. For example, how much would more violent crime prior arrests for Blacks than Whites affect the results? We will return to this issue shortly.

The comparisons between Table~\ref{tab:regwhites} and Table~\ref{tab:regblacks} illustrate three points. First, despite widespread claims of algorithmic bias, there is no manifest evidence of unfairness from this risk tool, at least for the machine learning method used, the provisional cost ratios imposed, the way the algorithm was tuned, and the conventional performance measures we computed. Racial bias in criminal justice risk assessments apparently is not inevitable. 

Second, should there be important racial differences, even if an artifact of how the algorithm was trained, those differences will get ported into practice. The training data and the algorithmic structure resulting from the training typically are not revisited. It can be very important, therefore, to examine the robustness of the results through several rounds of retuning coupled with new random splits of the data into training and test subsets. When this was done for these data, some racial differences surfaced. Some favored Black offenders, some favored White offenders, but none of the racial disparities would likely be consider large in practical terms. 

Third, results like those in Table~\ref{tab:regwhites} and Table~\ref{tab:regblacks} often will not convince all stakeholders. Empirical evidence of equitable outcomes does not necessarily carry the day because of strongly held beliefs that the entire criminal justice system is rife with racial bias. How can such a system and the data it assembles produce anything that is fair? Our empirical results could easily be dismissed as an aberration. From this perspective, the only reasonable action is to proactively reject all empirical risk assessments and rely instead of major structural and procedural reforms that achieve social justice. If we knew what kinds of structural and procedural reforms would secure full social justice, and if we were confident that these reforms could be implemented quickly and with integrity, the proactive rejection of all risk assessment might have some merit. In short, one side sees the glass as half full and aims to add a bit more water. The other side sees the glass completely empty and seeks to start again with a new glass and a new water supply.  

\subsection*{Training the Algorithm on White Offenders Only}

Perhaps there is a constructive middle ground. One can take the claims of racial unfairness seriously and try to respond to them as risk assessment tools are built. In particular, one can frame claims of racial bias affecting Black offenders as implying that the treatment of White offenders is a manifestation of ``White privilege.''  Algorithmic training can usefully respond to this view.

One starts by training the algorithm \textit{only on White offenders}. Risk classification is then undertaken separately for Black and White offenders using their own test data. With the training on only Whites completed, the two sets are test data can be used for predicting different risk classes. We emphasize the algorithmic structure arrived at with the White training data is used separately with the White and Black test data. Risks for Black and White offenders are processed by the trained algorithm as if everyone is White because that is all the algorithm ``knows.'' 

\begin{table}[htp]
\footnotesize
\caption{Stochastic Gradient Boosting Confusion Table from Test Data for White Offenders Using the White-Trained Algorithm: 59\% Predicted No Arrest, 32\% Predicted Non-violent Arrest, 9\% Predicted Violent Arrest}
\begin{center}
\begin{tabular}{|c|c|c|c|c|}
\hline \hline
Observed & No Arrest  & Non-Violent Arrest & Violent Arrest & Classification \\
Outcomes & Predicted &  Predicted & Predicted &  Error \\ \hline
No Arrest  & 24773 & 8030 & 2520 & .30 \\
Non-Violent Arrest & 9063 & 9520 & 2203 & .54 \\
Violent Arrest & 1841 & 1541 & 1033 & .77 \\
\hline
Prediction Error & .30 & .51 & .82 & \\
\hline \hline 
\end{tabular}
\end{center}
\label{tab:whites}
\end{table}

Table~\ref{tab:whites} is the confusion table for White offenders. The results are very similar to those in Table~\ref{tab:regwhites}. For example, 59\% are predicted to not be re-arrested, 32\% are forecasted be re-arrested for a non-violent crime, and 9\% are predicted to be re-arrested for a violent crime. This does not differ in important ways from the predicted risk classes when the algorithm was trained on all of the data. One might expect that accuracy would be better in Table~\ref{tab:whites} because the training data and test data are for White offenders. One possible implication is that the social and law enforcement processes responsible for recidivism are very similar for both racial groups. But we need to dig deeper.

\begin{figure}[htbp]
\begin{center}
\includegraphics[width=3.5in]{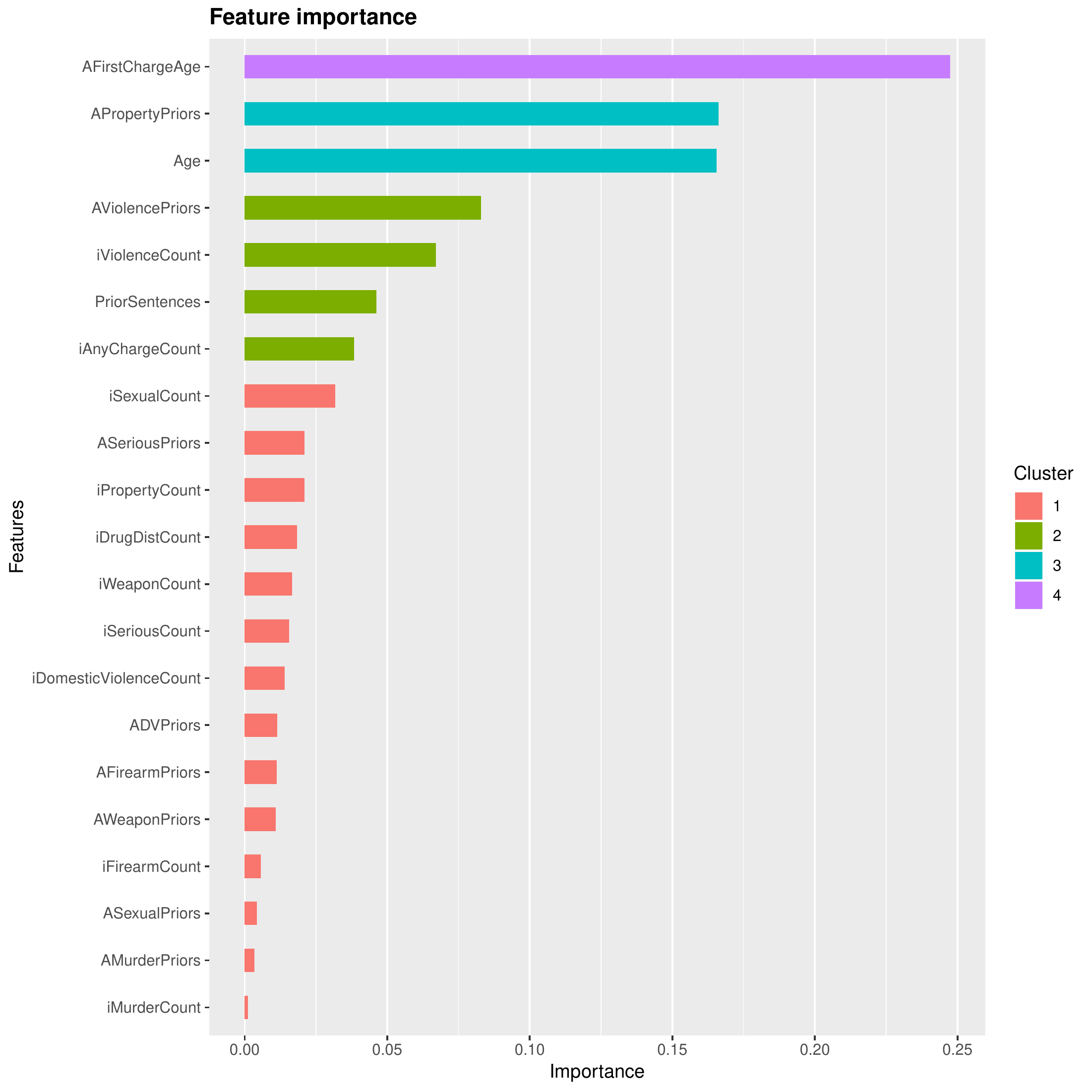}
\caption{Input Contributions To Stochastic Gradient Boosting Fit
for White Offenders Trained on White Offenders 
(contribution sum to 100\%)}
\label{fig:imp1}
\end{center}
\end{figure}

Figure~\ref{fig:imp1} shows for White offenders the fitting importance of each predictor. The prefix ``A'' denotes priors from arrests as an adult. The prefix ``i'' denotes charges from the instant crime for which an offender is being arraigned. Priors and counts are integers. Each predictor's contribution to the fit is computed for each pass through the data. Importance is computed as the average over passes, standardized so that the average contributions over all predictors sum to 100\%. The formal rationale can be found in Hastie et al., (2009 section 15.3.2). 

\begin{figure}[htbp]
\begin{center}
\includegraphics[width=2.5in]{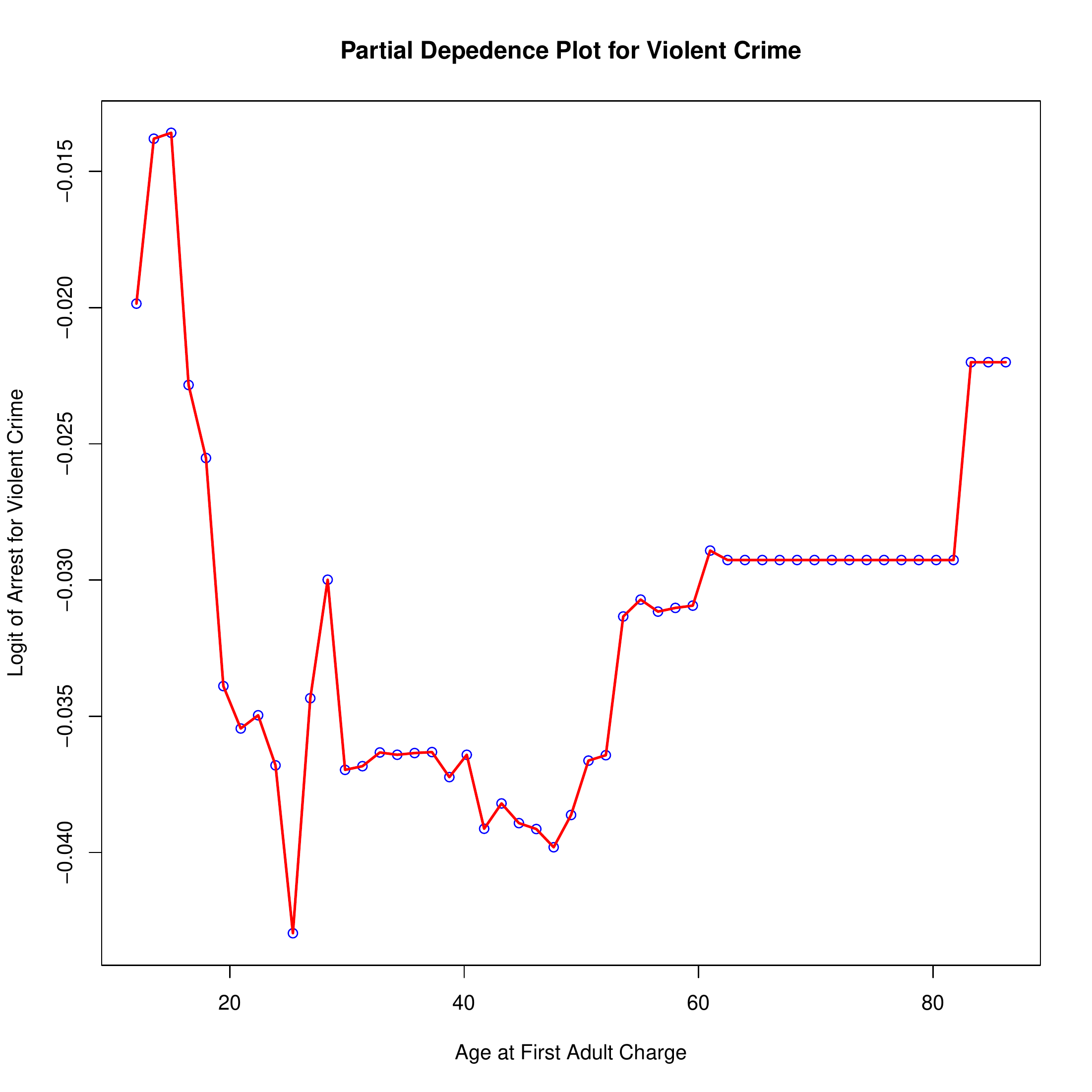}
\caption{Partial Dependence Plot of the Association of Age at the First Adult Charge with An Arrest for a Violent Crime for White Offenders Using A White-Trained Trained Algorithm}
\label{fig:first}
\end{center}
\end{figure}

\begin{figure}[htbp]
\begin{center}
\includegraphics[width=2.5in]{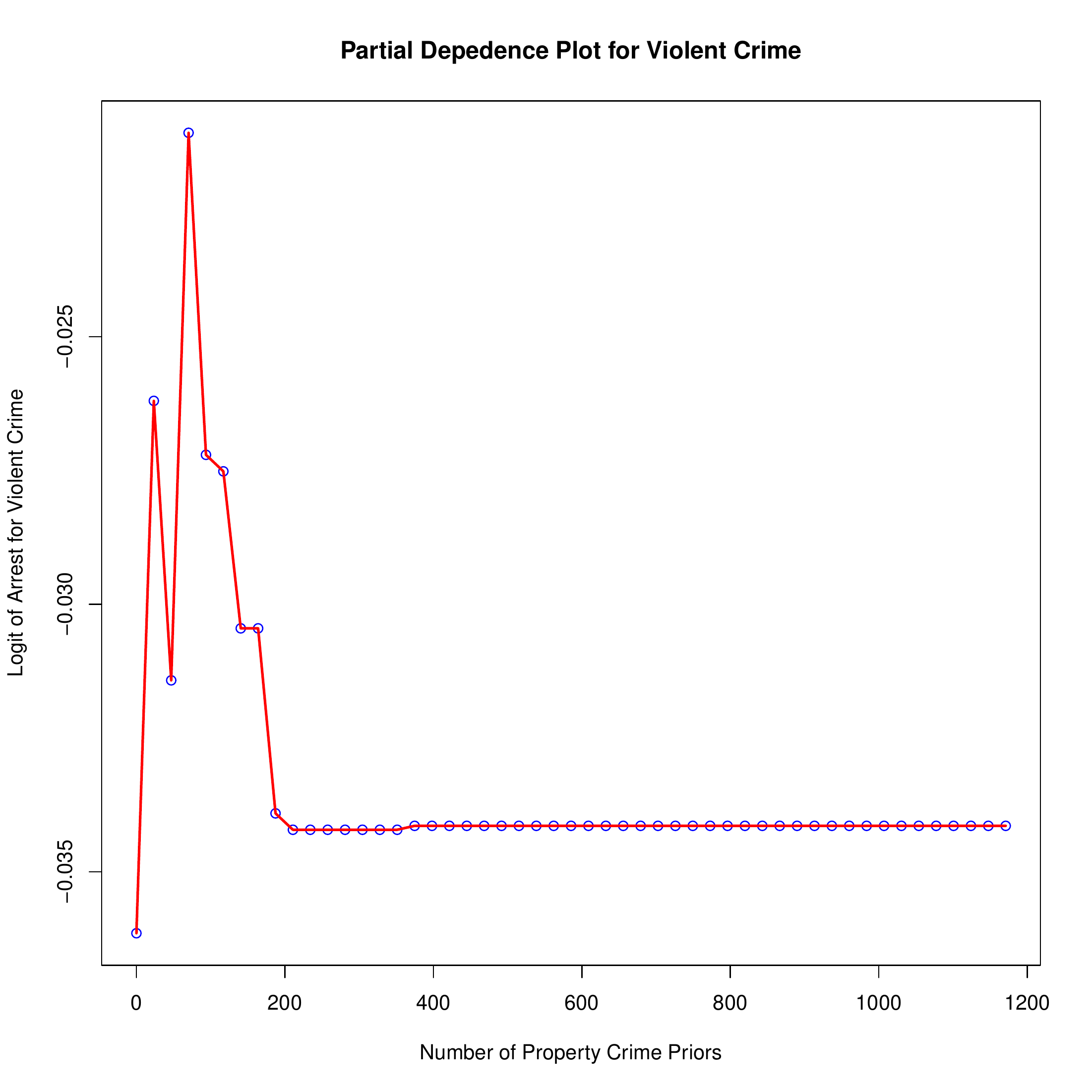}
\caption{Partial Dependence Plot of the Association of The Number of Priors for Property Crimes with An Arrest for a Violent Crime for White Offenders Using A White-Trained Trained Algorithm}
\label{fig:prop}
\end{center}
\end{figure}

\begin{figure}[htbp]
\begin{center}
\includegraphics[width=2.5in]{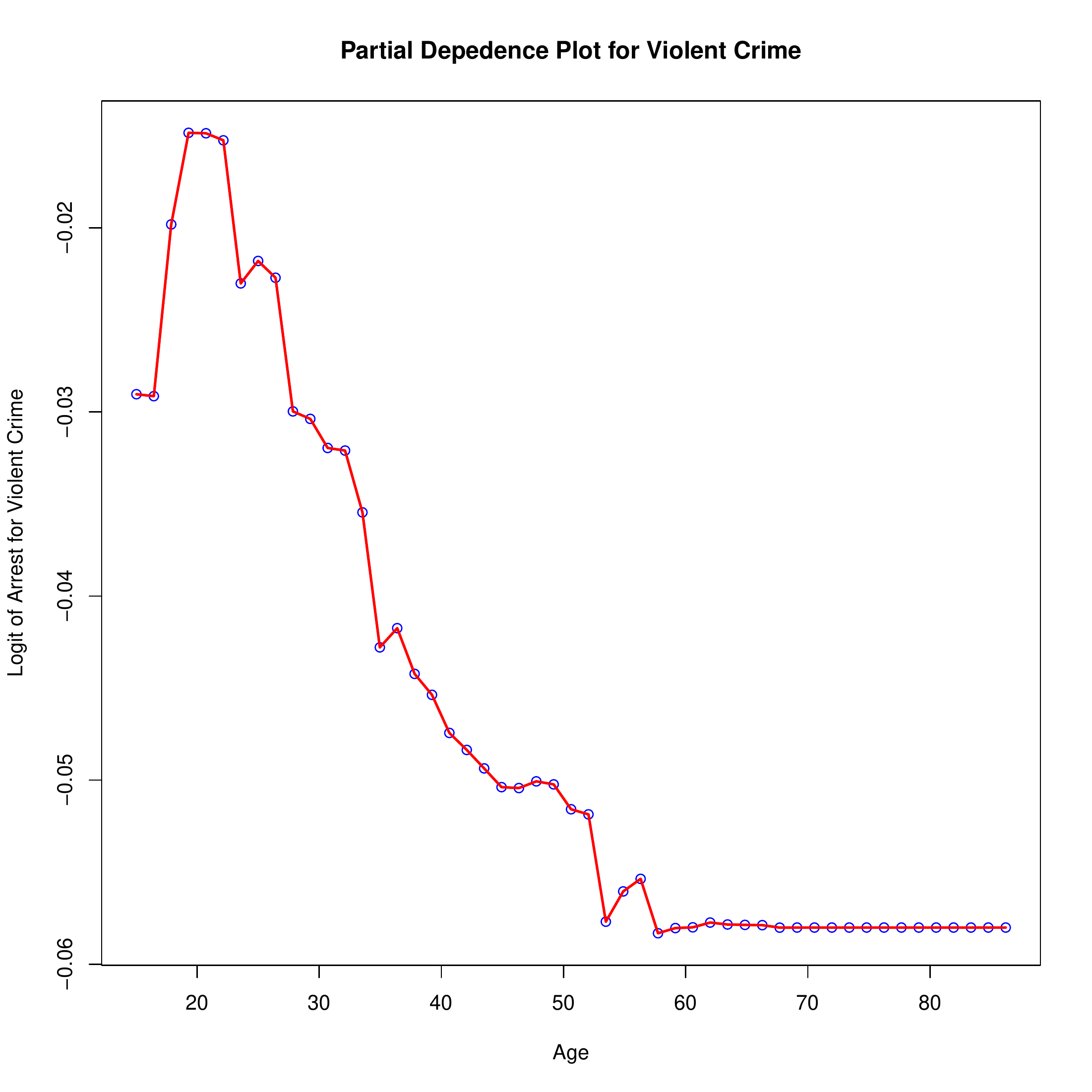}
\caption{Partial Dependence Plot of the Association of Age with An Arrest for a Violent Crime
for White Offenders Using A White-Trained Trained Algorithm}
\label{fig:age}
\end{center}
\end{figure}

The following three predictors, in order, dominate the fitting process. They will have important implications later. 
\begin{enumerate}
\item
\textbf{The age at which an offender is first charged \textit{as an adult}}. For a juvenile to be charged as an adult requires that the crimes responsible be very serious and typically violent as well. Not surprisingly, a very early adult charge is a powerful indication of subsequent crime (Berk 2017). But, the relationship with future arrests is non-linear. Computed risks decline sharply until the first adult change falls the early 20s, levels off (Berk, 2018), and then increases somewhat in middle age. The increase in middle age is thought to be related to various kinds of family violence. Figure~\ref{fig:first} is the partial dependence plot. We used the library \textit{pdp} in R. Logits (i.e., logged odds) are on the vertical axis. The formalities are addressed in Hastie et al. (2009: section 13.10.2).\footnote
{
The lines connecting the open circles are an interpolation, not a smoother. The sequence of fitted values after the age of 60 is the product of almost no data. To reduce computation burdens, the predictor values are binned.
}
\item
\textbf{The number of arrests for property crimes (in which no force is used)}. Its relationship with future arrests is thought to be positive. Figure~\ref{tab:whites} constructed using the same software shows the partial dependence plot in which the relationship is strongly positive where the mass of the training data are, and then turns negative for the very few case with more than 100 property priors. There are almost no cases in the training data with more than 200 property crime priors, and we suspect those numbers were recorded in error.\footnote
{
The issue is sparse \textit{training} data for cases with very large numbers of property priors. There is almost nothing to train on. The manner in which partial dependence plots are constructed uses the full dataset for each fitted value that is plotted; that's not the problem. To learn more details of how the plotting is done when there is a large number of predictor values, the documentation for \textit{partial} in the \textit{pdp} library should be consulted. For Table~\ref{tab:whites}, the binned, plotted points are 25 priors apart. 
}
The difference between no property priors and 25 property priors makes an important difference. Additional property priors make little difference.
\item
\textbf{The age of the offender.} Age also is well-known to be a powerful predictor, which has its peak impact in the late teens and early 20s, drops off sharply until about age 40, and then levels off (Berk, 2018). Figure~\ref{fig:age} is the partial dependence plot for age.\footnote
{
Data for individuals over 60 years of age is very sparse.
} 
\end{enumerate}

\begin{table}[htp]
\footnotesize
\caption{Stochastic Gradient Boosting Confusion Table form Test Data for Black Offenders Using the White-Trained Algorithm: 54\% Predicted No Arrest, 30\% Non-violent Arrest, 16\% Predicted Violent Arrest }
\begin{center}
\begin{tabular}{|c|c|c|c|c|}
\hline \hline
Observed & No Arrest  & Non-Violent Arrest & Violent Arrest & Classification \\
Outcomes & Predicted &  Predicted & Predicted &  Error \\ \hline
No Arrest  & 43167 & 16231 & 8353 & .36 \\
Non-Violent Arrest & 16488 & 15917 & 6739 & .60 \\
Violent Arrest & 5246 & 4563 & 3762  & .72\\
\hline
Prediction Error & .33 & .57 & .80 & \\
\hline \hline 
\end{tabular}
\end{center}
\label{tab:blacks}
\end{table}

Table~\ref{tab:blacks} is the test data confusion table for Black offenders constructed from the algorithm trained on White offenders; White privilege is available to White and Black offenders alike. Black offenders are a bit more likely than White offenders to be predicted to be re-arrested for a violent crime: 16\% to 9\%. Black offenders are a bit less likely than white offenders to be predicted to experience no re-arrest whatsoever:  54\% to 59\%. (The percentages for a re-arrest for a non-violent crime are far more alike.) Whether these differences represent important inequality is in some sense in the eyes of the beholder. But the over-representation of Black offenders in the violent crime class probably will be taken by some as evidence of ``bias.''\footnote
{
Proper statistical inference for stochastic gradient boosting has not yet been solved because the algorithm is adaptive, and one must take into account not just all of the regression trees there were actually construct by the boosting algorithm, but all of the regression trees that \textit{could} have been constructed (Berk et al., 2014). However, if one is prepared to treat the training data and the trained algorithmic structure as fixed, one can with test data compute legitimate confidence intervals and statistical tests using a non-parametric bootstrap. Because the number of test observations is very large, one easily rejects the null hypothesis of no difference at less than the .001 level for all of the White/Black comparisons. Given all of the caveats, statistical tests probably are not instructive. 
}

For two of the three outcomes, there is a little more forecasting error for Black offenders than White offenders: 33\% to 30\%, 57\% to  51\%, and 80\% to 82\%. Such comparison reveal virtually no differences because of tuning approximations. In this case, the direction the three comparisons was very stable, but the differences shown are on the high side. Still, some stakeholder could object. Further adjustment need to be considered. 

\subsection*{Discounting Prior Arrests}

The disparity between Black offenders and White offenders in the proportion forecasted to be re-arrested for a violent crime is large enough to explore further. We sought, therefore, to construct results that might be more politically acceptable by discounting for Black offenders' priors for serious crimes when the risks for a re-arrest were computed. The algorithm itself was trained on white offenders only.

Some might claim that the number of prior arrests for Black offenders, even for felonies and violent crimes, is too large by a factor perhaps as large as  2. Dividing the number of priors in the test data is a linear transformation that should make almost no difference in the results, especially because the impact on risks projected depend substantially on the first few priors. Therefore, with no theoretical or policy guidance, we employed for Black offenders the square root of the number of test data priors for the variety of serious priors included as predictors. We expected that this non-linear transformation would produce a somewhat different confusion table.

Table~\ref{tab:blacksadj} shows the confusion table that results. As before, risks for Black offenders are computed when test data for Blacks are used with the white-trained algorithm. We see that the square root adjustment to the numbers of priors does not improve fairness over the measures used. In particular, prediction error had increased for all three outcome classes. Probably the most important change is that the proportion of those projected to be re-arrested for a violent crime has increased from 16\% to 21\%. The gap between Black offenders and White offenders has increased substantially. 

Using a nonlinear transformation is perhaps the culprit. The algorithm was trained on the number of priors. The test data now employ a non-linear transformation of those numbers. The likely result is a reduction in performance. Perhaps some other transformation of the priors would perform better.\footnote
{
We also tried dividing the number of priors by 2, and as expected, almost nothing changed compared to Table~\ref{tab:blacks}. We also tried dividing by 2 and then recoding prior counts of 1 to 0  (i.e., turning many Blacks in first offenders, another non-linear transformation). That changed the confusion table in a manner much like Table~\ref{tab:blacksadj}. Fairness was not improved.
}

\begin{table}[htp]
\footnotesize
\caption{Adjusted Gradient Boosting Confusion Table from Test Data for Black Offenders Using the White-Trained Algorithm: 53\% Predicted No Arrest, 26\% Non-violent Arrest, 21\% Predicted Violent Arrest}
\begin{center}
\begin{tabular}{|c|c|c|c|c|}
\hline \hline
Observed & No Arrest  & Non-Violent Arrest & Violent Arrest & Classification \\
Outcomes & Predicted &  Predicted & Predicted &  Error \\ \hline
No Arrest  & 41927 & 15182 & 11841 & .30 \\
Non-Violent Arrest & 17588 & 12100 & 9414 & .70 \\
Violent Arrest & 5485 & 3800 & 4129 & .69\\
\hline
Prediction Error & .36 & .61 & .84 & \\
\hline \hline 
\end{tabular}
\end{center}
\label{tab:blacksadj}
\end{table}

\subsection*{Altering the Base Rates}

Another reason why Black offenders might have greater computed risks for violent crime re-arrests is base rate differences between Blacks and Whites. Recall that in Table~\ref{tab:whites}, 9\% of White offenders were in fact re-arrested for a violent crime after an arraignment release, and in Table~\ref{tab:blacks}, 16\% of Black offenders were in fact re-arrested for a violent crime after an arraignment release. Differences in base rates are known to cascade through a confusion table potentially producing several different kinds of unfairness (Kleinberg et al., 2017).\footnote
{
Black base rate for violent crimes may well be an under-estimate because police clearance rates in disadvantaged neighborhoods are well known to generally be lower than in other neighborhoods, even for crimes like homicide (Lowery, 2019). Part of the explanation is that the kinds of crimes and the attributes of perpetrators create more challenges in some neighborhoods than others. For example, in some neighborhoods potential witnesses are more likely to fear for their safety and be less inclined to come forward. The mix of crimes matters too. Homicides associated with intimate partner violence, for instance, automatically define ``a person of interest'' who usually is easily found. Drive-by shootings usually are more challenging. 
}

It is easy to change base rates for re-arrests for violent crime. One merely re-weights the data so that new arrests of Black offenders for violent crime are effectively reduced relative to new arrests of White offenders for violent crime. We have undertaken such exercises elsewhere (Berk, 2019; Elzarka, 2019) showing that changing bases rates can dramatically alter the results and favorably impact fairness. 

In this case, we ran the stochastic gradient boosting algorithm again on whites alone, but used weighting to discount the importance of re-arrests for violent crime. Consequently, the algorithm did not work nearly as hard to fit re-arrests for a violent crimes, regardless of the background of the offender. That is, for all offenders, those who in the white training data who were more likely to be re-arrested for a violent crime, got a break. We anticipated that because Black offenders empirically were more likely than White offenders to be re-arrested for violent crimes, the violent re-arrest base rates for the two groups implicitly would become more alike. Then, when risks were computed for Blacks and Whites using their test data, the forecasts of violent crime re-arrests would be similar.

Table~\ref{tab:whitesdown} is the resulting confusion table for White offenders constructed from test data. Table~\ref{tab:blacksdown} is the resulting confusion table for Black offenders constructed from test data. No adjustments were made, as we did before, for the numbers of priors for more serious crimes; interest centered on the impact adjusted base rates. 

\begin{table}[htp]
\footnotesize
\caption{Stochastic Gradient Boosting Confusion Table from Test Data for White Offenders With Violent Crime Re-Arrests Down-weighted Using White-Trained Algorithm: 58\% Predicted No Arrest, 42\% Non-violent Arrest, 3\% Predicted Violent Arrest }
\begin{center}
\begin{tabular}{|c|c|c|c|c|}
\hline \hline
Observed & No Arrest  & Non-Violent Arrest & Violent Arrest & Classification \\
Outcomes & Predicted &  Predicted & Predicted &  Error \\ \hline
No Arrest  & 23882 & 11027 & 719 & .33 \\
Non-Violent Arrest & 8131 & 11915 & 449 & .42 \\
Violent Arrest & 1687 & 2209 & 375 & .92\\
\hline
Prediction Error & .30 & .53 & .77 & \\
\hline \hline 
\end{tabular}
\end{center}
\label{tab:whitesdown}
\end{table}

 \begin{table}[htp]
\footnotesize
\caption{Stochastic Gradient Boosting Confusion Table from Test Data for Black Offenders With Violent Crime Re-Arrests Down-weighted Using White-Trained Algorithm: 51\% Predicted No Arrest, 44\% Non-violent Arrest, 5\% Predicted Violent Arrest }
\begin{center}
\begin{tabular}{|c|c|c|c|c|}
\hline \hline
Observed & No Arrest  & Non-Violent Arrest & Violent Arrest & Classification \\
Outcomes & Predicted &  Predicted & Predicted &  Error \\ \hline
No Arrest  & 41371 & 23664 & 2836 & .39 \\
Non-Violent Arrest & 15182 & 21865 & 2061 & .44 \\
Violent Arrest & 5315 & 6893 & 1279 & .91\\
\hline
Prediction Error & .34 & .58 & .79 & \\
\hline \hline 
\end{tabular}
\end{center}
\label{tab:blacksdown}
\end{table}

The differences between the computed risk distributions are much smaller and now perhaps of no policy importance. For White offenders compared to Black offenders and from no re-arrest to a re-arrest for a violent crime one has: 58\% to 51\%, 42\% to 44\%, and most important, 3\% to 5\%. Forecasting error is also rather similar: 30\% to 34\%, 53\% to 58\%, and 77\% to 79\%. 

By these criteria, the results may well be more politically acceptable than the original results in which both White and Black offenders had their risk scores computed from test data using the white-trained algorithm. Now, all offenders are sentenced is if they were white, conditional on White and Black offenders having more similar bases rates for violent crime re-arrests. 

The further tuning and different cost ratios, it would be possible to do even better. We doubt, however,  that perfect equality is possible in the real world of criminal justice practice. But once again, the baseline is current practice. The goal is improvement.\footnote
{
If one is prepared to leave behind the real world of criminal justice practice, perfect equality is easily obtained. One can simply flip a coin to determine who is released at arraignment. This is absolutely fair because all offenders, whatever their background, have the exact same chance of being released. The price is lots of mistakes because by chance many low risk offenders will be detained and many high risk offenders will be released. One has given up on accuracy. And if one gives up on accuracy, two other perfectly fair approaches are to release no one or to release everyone. 
}
  
\section*{Summary and Conclusions}

We began with conventional practice. The boosting algorithm was trained on Black and White offenders together, and possible racial differences were examined using test data separately for Black offenders and White offenders. There was no compelling evidence for racial unfairness and accuracy was improved compared to predictions from the marginal distribution of the response. Unfairness is apparently not inevitable. As  political matter, however, the empirical results might not carry the day.

We then illustrated with real data analysis methods by which one can arrive at risk assessment results that perhaps can be politically acceptable. We began by training a stochastic gradient boosting algorithm only on White offenders. Risk scores were then computed with test data separately from Black offenders and White offenders using the white-trained algorithm. One might argue that as a conceptual matter, treating Black offenders as if there were White provides some protection again ``biased'' algorithms. 

Some might find those results sufficiently fair. Others would be troubled by the greater fraction of Black offenders than White offenders who were forecasted to be re-arrested for a violent crime. One possible explanation was that because of ``biased'' policing or ``over-policing,'' Blacks offenders undeservedly had longer prior records. Those longer records increased the computed risks for re-arrests for crimes of violence. 

We addressed those concerns in three ways. First, we had earlier discarded priors for ``petty offenses'' in which police could exercise considerable discretion. They played no role in how the algorithm was trained for any of the analyses and cannot be blamed for the results. Second, beginning with the white-trained algorithm, we reduced for Black test data the remaining priors by working with their square roots. The non-linear transformation changed the confusion table dramatically, and the results were less fair. The algorithm was trained in counts of prior arrests. Risk was computed using a non-linear transformation of those counts. One should expect important differences in the results on statistical grounds alone.

Third, differences in base rates are well known be a potential source of unfairness in criminal justice applications. Using weighting, we discounted in the white training data all re-arrests for violent crime, anticipating that implicitly the base rates by race would become more alike. Black offenders would benefit more than White offenders when re-arrests for violent crime were made relatively less common for all offenders. The over-representation of Black offenders projected to be re-arrested for crimes of violence was substantially reduced, and the proportion of Black offenders projected to be re-arrested for a violent crime was more like the proportion of White offenders projected to be re-arrested for a violent crime.  

There are surely no guarantees that politically acceptable results will be obtained with other data. Perhaps our remarkably fair initial results were an aberration. Regardless, we illustrated four potential, remedial strategies for risk assessments initially criticized as unfair. 
\begin{enumerate}
\item
Exclude predictors that arguably are especially vulnerable to racial bias. Good candidates are prior arrests for crimes in which police can exercise wide discretion.
\item
Train the risk algorithm with training data from the most privileged group. Then, compute risk for members of all groups separately from test data using the results from the algorithm trained on the most privileged group. These two steps alone my yield sufficient fairness once predictors vulnerable to bias are excluded.
\item
Consider discounting the impact of priors in test data for the less privileged groups. But a sensible transformation must be determined. 
\item
Also, consider retraining that algorithm on data from the privileged group discounting re-arrest for crimes thought to foster unfair forecasts.  Then proceed as in \#1 and \#2.
\end{enumerate}

It is likely that risk assessments can be undertaken that are substantially more politically acceptable to key stakeholders. However, the impact of each of the four strategies also must be evaluated for their impact on \textit{victims}. Any approach that discounts prior record and/or re-arrests for certain kinds of offenses, discounts the harm to victims of those crimes. Often those victims will be disproportionately from disadvantaged neighborhoods. In 2018, there were over 351 homicides in the city from which our data were collected. 92\% of the victims were African-American (McSwan, 2019). Is it fair to discount their deaths? 
 
\section*{References}
\begin{description}
\item
Berk, R.A. (2018) \textit{Machine Learning Forecasts of Risk in Criminal Justice Settings}. New York: Springer.
\item
Berk, R.A. (2019) ``Accuracy and Fairness for Juvenile Justice Risks Assessments.'' \textit{Journal of Empirical Legal Studies}, published online February, 2019 (DOI: 10.1111/jels.12206)
\item
Berk, R.A., Barnes, G., Alhman, L., \& Kurtz, E. (2010) ``When Second Best Is Good Enough: A Comparison Between a True Experiment and a Regression Discontinuity Quasi-Experiment.'' \textit{Journal of  Experimental Criminology} 6(2): 191--208.
\item
Berk, R.A., Heirdari, H., Jabbari, S., Kearns, M., \& Roth, A. (2018) ``Fairness in Criminal Justice Risk Assessments: The State of the Art.'' \textit{Sociological Methods and Research}, first published July 2nd, 2018, http://journals.sagepub.com/doi/10.1177/0049124118782533.
\item
Blinder, A.S. (1973). ``Wage Discrimination: Reduced Form and Structural
Estimations.''\textit{Journal of Human Resources} 8: 436--455.
\item
Bloom, S. L. (2014). ``The Impact of Trauma on Development and Well-Being.'' In K. G. Ginsburg \& S. B. Kinsman (Eds.), \textit{Reaching Teens - Wisdom From Adolescent Medicine.} Elk Grove Village, IL: American Academy of Pediatrics.
\item
Elzarka, A. (2019) ``Establishing Fairness in Algorithms.'' Thesis in Data Science,
presented to the faculties of the University of Pennsylvania in partial
fulfillment of the requirements for the degree of master of science in
engineering.
\item
Harcourt, B.W. (2007) \textit{Against Prediction: Profiling, Policing, and Punishing in an Actuarial Age}. Chicago, University of Chicago Press.
\item
Hastie, T., Tibshirani, R., \& J. Friedman (2009) \textit{The Elements of Statistical Learning}. New York: Springer
\item
Huq, A.Z. (2019) ``Racial Equality in Algorithmic Criminal Justice.'' \textit{Duke Law Journal} 68 (6), 1043--1134. 
\item
Kearns, M., Neel, S., Rith, A, \& Wu, Z. (2018) Preventing Fairness Gerrymandering: Auditing and Learning Subgroup Fairness.'' arXiv:1711.05144v4 [cs.LG].
\item
Kitagawa, E. (1995)``Components of a Difference Between Two Rates.'' \textit{Journal of the American Statistical Association} 50: 1168--1194. 
\item
Kleinberg, J., Mullainathan, S., \& Raghavan, M. (2017) ``Inherent Tradeoffs in the Fair Determination of Risk Scores.'' Proc. 8th Conference on Innovations in Theoretical Computer Science (ITCS).
\item
"Lowery, W., Kelly, K., Melnick, T \& S.Rich (2018) ?Where Killings Go Unsolved.? Washington Post, June 6, 2018. https://www.washingtonpost.com/graphics/2018/in murders-go-unsolved/
\item
Mayson, S.G., (2019) ``Bias In, Bias Out.'' \textit{The Yale Law Journal} 128: 2218--2300.
\item
McSwain, W.M. (2019) ``Police Deserve DA's Support.'' Philadelphia Inquirer, May 12 12, 2019, page C4.
\item
Oaxaca, R. (1973). ``Male-Female Wage Deferential in Urban Labor Markets.''
\textit{International Economics Review} 14 (3): 693--709.
\item
Starr, S.B. (2014) ``Evidence-Based Sentencing and the Scientific Rationalization of Discrimination. \textit{Stanford Law Review} 66: 803 -- 872.
\end{description}
\end{document}